\documentclass{article}%
\usepackage{amsfonts}
\usepackage{amsmath}
\usepackage{amssymb}
\usepackage{graphicx}%
\begin{document}

\title{Stochastic Price Dynamics Equations Via Supply and Demand; Implications for
Volatility and Risk}
\author{Carey Caginalp\\Economic Science Institute, Chapman University, Orange, CA
\and Gunduz Caginalp\\Mathematics Department, University of Pittsburgh, Pittsburgh, PA }

\begin{center}
\textbf{Derivation of non-classical stochastic price dynamics equations}

\bigskip

Carey Caginalp$^{1,2}$ and Gunduz Caginalp$^{1}$

\bigskip

July 20, 2020 2 pm
\end{center}

\bigskip

\bigskip

$^{1}$ Mathematics Department, University of Pittsburgh, Pittsburgh, PA 15260

$^{2}$ Economic Science Institute, Chapman University, Orange, CA 92866

\bigskip

\textbf{E-mail:} CC \ carey\_caginalp@alumni.brown.edu, GC \ caginalp@pitt.edu\ 

\bigskip


\bigskip

\textbf{Abstract.} We analyze the relative price change of assets starting
from basic supply/demand considerations subject to arbitrary motivations. The
resulting stochastic differential equation has coefficients that are functions
of supply and demand. We derive these rigorously. The variance in the relative
price change is then also dependent on the supply and demand, and is closely
connected to the expected return. An important consequence for risk assessment
and options pricing is the implication that variance is highest when the
magnitude of price change is greatest, and lowest near market extrema. This
occurs even if supply and demand are not dependent on price trend. The
stochastic equation differs from the standard equation in mathematical finance
in which the expected return and variance are decoupled. The methodology has
implications for the basic framework for risk assessment, suggesting that
volatility should be measured in the context of regimes of price change. The
model we propose shows how investors are often misled by the apparent calm of
markets near a market peak. Risk assessment methods utilizing volatility can
be improved using this formulation.

\bigskip


\bigskip


\bigskip

\pagebreak

\textbf{1. Introduction}. While equilibrium price theory has been developed
extensively in classical economics, the study of dynamics that describes the
path to equilibrium is still in the developing stage. Broadly speaking, the
approaches can be divided into (i) discrete models -- based on an imbalance of
supply and demand -- that have typically been considered for goods and
commodities, (ii) continuum models -- often incorporating randomness -- for
asset price dynamics that are standard in options pricing and risk assessment.

\bigskip

\textbf{1.1.} A simple and standard model of type (i) above is the excess
demand model, which is often stated in classical economics\footnote{Walrasian
equilibrium is defined by zero excess demand, $E=D-S,$ i.e., the right hand
side of $\left(  \ref{discrete p}\right)  ,$ vanishes so that the price is
unchanged in the next time period. The price thereby adjusts to the quantity.
A tatonnement process (see e.g., \cite{WE}), such as $\left(  \ref{discrete p}%
\right)  $ is one in which trading occurs only at equilibrium. When trading at
the discrete time $t$ has ended, the intersection of the new supply and demand
then determines the price $P_{t+1}$ of the next trade.} as
\begin{equation}
p_{t}-p_{t-1}=\frac{1}{\tau_{0}}\left(  d_{t-1}-s_{t-1}\right)
\label{discrete p}%
\end{equation}
where $p_{t}$ is the price at the discrete time, $t$, and $s_{t-1}$ and
$d_{t-1}$ are the supply and demand at time $t-1$, and $\tau_{0}^{-1}$ is a
constant that determines the extent to which prices move for each unit of
imbalance between supply and demand (e.g. Watson and Getz, 1981 \cite{WG} or
Weintraub 1979 \cite{WE}). This is obtained directly from supply and demand
functions that are assumed to be straight lines, which are good approximations
for small deviations when these functions are smooth. Of course, $\left(
\ref{discrete p}\right)  $ is only a local equation that is valid for a
particular pair of linear supply and demand. For example, an imbalance created
by $d_{t-1}=10,020$ and $s_{t-1}=10,000$ will have a much smaller impact on
price change than would $d_{t-1}=40$ and $s_{t-1}=20.$ This demonstrates the
need for normalization, realized by dividing the right hand side of $\left(
\ref{discrete p}\right)  $ by $s_{t-1}.$ Similarly, the left hand side must be
normalized by dividing by $p_{t-1}$ leading to the equation%
\begin{equation}
\frac{p_{t}-p_{t-1}}{p_{t-1}}=\frac{1}{\tau_{0}}\frac{d_{t-1}-s_{t-1}}%
{s_{t-1}}. \label{lin discrete}%
\end{equation}

While these normalizations lead to an equation that is a reasonable non-local
model, another feature of $\left(  \ref{discrete p}\right)  $ is that it is a
linear equation, so that the price change is always proportional to the excess
demand. While linearity is often a convenient and reasonable approximation,
there is no compelling requirement that price change be a linear function of
excess demand. Introducing a differentiable function $g:\mathbb{R}%
^{+}\mathbb{\rightarrow R}$ with suitable properties including $g\left(
1\right)  =0$ and $g^{\prime}>0,$ we can write in place of $\left(
\ref{discrete p}\right)  $ the equation%
\begin{equation}
\tau_{0}\frac{p_{t}-p_{t-1}}{p_{t-1}}=g\left(  \frac{d_{t-1}}{s_{t-1}}\right)
. \label{nonlin discrete p}%
\end{equation}

Thus, information on the supply and demand at a any discrete time determines
the price change for the next discrete time. The design of markets and
efficient price discovery has been an active research area from both a
theoretical and experimental perspective. See models \ in Milgrom, 2017
\cite{MI} and Gjerstad and Dickhaut, 1998 \cite{GD}, Hirshlefer et. al., 2005
\cite{HG}, Gjerstad, 2007 \cite{G} and 2013 \cite{G1} and references therein.
The experimental aspect has been studied by researchers including Plott and
Pogorelskiy, 2017 \cite{PP}, Crocket et. al., 2009 \cite{GD}, Bossaerts and
Plott, 2004 \cite{BP}, Porter and Rassenti, 2003 $\cite{PR}.$

\bigskip

\textbf{1.2.} At the continuum level, the focus of research for several
decades has been on the price dynamics of asset prices, such as stocks and
options, subject to randomness. A standard equation that has been the starting
point for much of mathematical finance is written in terms of Brownian motion
$W\left(  t;\omega\right)  $, with $\omega\in\Omega$, the probability state
space, as
\begin{equation}
P^{-1}\frac{dP}{dt}=\mu dt+\sigma dW. \label{BS}%
\end{equation}
Here $P$ is the price as a function of continuous time, $t$, while $\mu$ and
$\sigma$ are the expected return and standard deviation. The parameter $\mu$
is the expected return (often based on historical data), and $\sigma$ is often
set based on the idea (that has some limited empirical justification) that
volatility remains relatively stable in time. \ These parameters are often
assumed to be constant, and in some cases a prescribed function of time. This
equation has a long history, with the main ideas dating back to Bachelier,
1900 \cite{BA} and now utilized in basic textbooks such as Karatzas and
Shreve, 1998 \cite{KS} and Wilmott, 2013 \cite{W}. With no information besides
some price history, one can regard $\left(  \ref{BS}\right)  $ as a good first
approximation for several mathematical problems such as options pricing and
risk assessment.

The approach leading to $\left(  \ref{BS}\right)  $ does not build on the
basic economic ideas of supply and demand, but rather has limited
justification based on empirical observation and is appealing due to the
salient mathematical properties. One of the deficiencies of $\left(
\ref{BS}\right)  $ is that it vastly understates the risk for unusual events.
For example, if one calculates the probability of, say, a $4\%$ or $5\%$ drop
in a stock index based upon the empirical daily standard deviation, one
obtains a result that is vastly smaller than empirical observations. While
many explanations (with some justification) have been offered for this
anomaly, often called "fat tails," Caginalp and Caginalp 2018, \cite{CC},
\ 2019, \cite{CC1} showed that it arises as a consequence of the mathematical
property that a quotient of normals is approximately normal in part of the
domain of the density (see D\'{\i}az-Franc\'{e}s and Rubio 2013 \cite{DR},
Champagnat et. al. 2013 \cite{CH} and references therein), but not near the
tail, where it can be a power law rather than exponential. The basic idea is
that supply and demand can be expected to be normal as a consequence of the
Central Limit Theorem. Thus, their quotient as one has in $\left(
\ref{lin discrete}\right)  $ or $\left(  \ref{nonlin discrete p}\right)  $
will \textbf{not} be close to normal near the tail of the distribution, i.e.,
for rare events.

A practical examination of the source of randomness in asset prices shows that
the overwhelming fraction of the randomness arises from the randomness in
supply and demand. In other words, if one knew how the supply and demand would
change at the next time interval, one would essentially know how prices would
evolve\footnote{One can almost regard this statement as a tautology since the
price is determined exclusively by the supply and demand functions.
Theoretically,\ the price change after a small time increment will be
determined by the change in these functions. In many markets there are
features that complicate the supply/demand analysis, such as market makers who
can buy on their own account at better prices than the traders.}. \ In trading
of a major stock or index, for example, there are dozens of professionals
focusing on the same stock and whose sole business consists of exploiting very
small deviations from optimal pricing. Indeed, these professionals observe the
same shifts in supply and demand again and again, and earn their living by
adjusting their bids and asks in response to these shifts.

\bigskip

These professionals, some of whom are "market makers"\ tasked with ensuring
orderly markets, are not concerned with the fundamentals or even the long term
trends. At any given moment they observe the change in the supply demand
curves regardless of origin of those changes. Given the change in supply and
demand, there will be a unique price that is clearly based on previous
iterations. If some of the professionals have biases or irrational
expectations whereby they are not able to deduce the correct new price, they
will not be in business very long since there are many trades throughout one
day, so that a losing strategy will exhaust the capital in a short time. This
does not contradict the premises of behavioral finance since there are many
aspects of trading, with some having an efficiency higher than others. The
perspective of the market makers is that they have some capital for their
business which involves adjusting to the orders of a stock. Unlike individual
or hedge fund investors or even mutual fund managers, the market makers cannot
tie up their business capital based on their assessment of either the
fundamental value or the long term trend (and other technical indicators), or
other beliefs they may hold about the stock.

The situation is illustrated in Figure 1. The supply-demand graph displays the
initial supply function (which is increasing) in solid red\ and the
(decreasing) demand function in solid blue. The price is established by the
intersection at the price $P=1.29.$ A short time later, there is a random
event (e.g., a news announcement such as an earnings report) that changes the
supply and demand for the asset. In particular, if the news is positive, it
will increase the demand, so the solid blue curve shifts upward to the dotted
blue line. The potential sellers are also aware of the positive news and
re-adjust their orders to reflect the changing circumstances, thereby raising
the price that they are willing to accept. The solid red curve thus shifts
upward to the dotted red curve. The new supply and demand functions now
intersect at a higher price, $P=1.44.$ The key point is that the random
event$\allowbreak$ influences a large number of agents in terms of their
preferences to buy or sell, so that the impact of the randomness is entirely
comprised of the shifts in the supply and demand curves. Given the shifts in
the supply and demand curves, there is little additional source of randomness
given the large number of market makers who are seeking to capitalize on these
shifts and have optimized on the same shift many times.%

\begin{center}
\fbox{\includegraphics[
natheight=2.008100in,
natwidth=3.512900in,
height=2.0081in,
width=3.5129in
]%
{D:/ResearchStochasticsAssetFlowFourthPaper/PhysicaA-submission-Derivation/graphics/ManuscriptJuly20-2020RevisionSubmitPublisher__1.png}%
}\\
Figure 1. The supply-demand graph displays the initial supply function (which
is increasing) in solid red and the (decreasing) demand function in solid
blue. The price is established by the intersection at the price $P=1.29$. A
short time later a random event increases demand and decreases supppy, so the
blue curve shifts upward to the dotted blue line, and the supply shifts
downward to the dotted red line, establishing a higher price, $P=1.44$.
\label{SD}%
\end{center}

\bigskip

Examples of company specific news\ that influences the supply/demand curves
include quarterly revenue and earnings reports, changes in the company's
leadership, securing a lucrative contract, announcements on restating previous
earnings, government announcements of investigations into the company, etc.
For the broader market, the supply/demand curves often shift with government
updated indicators, most of them monthly, such as the nonfarm payrolls,
stating the net number of jobs added or lost during the preceding month, the
retail sales changes, the consumer price increases, trade balance numbers,
etc. Other factors are also include changes in interest rates, i.e., the bond
markets, international trade and currencies, natural disasters, etc. Large
changes in supply/demand occur when there is large deviation from the expected
outcome. For example, during both the financial crisis of 2008 and the Covid
pandemic of 2020, the employment announcements were very significant as the
markets braced for job loss of hundreds of thousands or millions,
respectively. What is often surprising to non-experts, however, is that
supply/demand curves prior to an announcement are based on the current
forecast. If that forecast is for two million jobs lost, and the news is that
"only" one million jobs are lost, this is favorable news, and the supply (of
shares submitted for sale in the S\&P index, for example) often shifts down,
and the demand shifts up (as displayed in Figure 1). Thus, the impact of the
news is always relative to the existing expectations.

This idea can be tested empirically in exchanges and experimentally in
laboratories. Given the same shift in supply and demand how much variance will
there be in a market with many experienced market makers and short term
traders? We claim that it will be negligible compared to the randomness in
supply and demand arising from news items (e.g., earnings reports, forecasts
and analysts reports for a stock) and influx or outflow of funds for a
particular asset.

\bigskip

In particular, one can design a market in which news (altering the payout)
will arrive at various times. In addition to the usual traders, there will be
short term traders who have the constraint that they must have zero inventory
of the asset at the end of each period (similar to a market maker). By
analyzing the supply/demand changes throughout the experiment, one can
determine the variance in the relative price change among the times when the
shift in supply/demand is nearly identical. The hypothesis is that when the
shift in supply/demand is similar, so is the price change. On the other hand,
one can determine the variance in the supply/demand shift given a spectrum of
news that impacts the asset payoff.

\bigskip

\bigskip

Thus, a fundamental analysis of randomness in asset prices should begin with
an examination of the process by which randomness in supply and demand
propagates to the stochastics of price change.

Our goal here is to present a precise derivation and justification of an
equation analogous to $\left(  \ref{BS}\right)  $ that is based on supply and
demand considerations (see Caginalp and Caginalp, 2019, \cite{CC2}), namely,%
\begin{equation}
P^{-1}dP=G\left(  D/S\right)  dt+\sigma\frac{D}{S}G^{\prime}\left(
D/S\right)  dW. \label{CC}%
\end{equation}
Here, $S$ and $D$ be the expected value of supply and demand, respectively.
The basic premise is that the relative change in price, $P^{-1}dP/dt$ in terms
of a function $G$, which is analogous to $g$ above, and meets the requirements
specified in Section 2. Since $G$ depends on the ratio of total supply,
$\tilde{S},$ and demand, $\tilde{D}$, one has the basic equation
$P^{-1}dP/dt=G\left(  \tilde{D}/\tilde{S}\right)  .$ Writing%
\[
\tilde{D}\left(  t,\omega\right)  =D\left(  t\right)  \left(  1+\frac{\sigma
}{2}R\right)  ,\ \ \tilde{S}\left(  t,\omega\right)  =S\left(  t\right)
\left(  1-\frac{\sigma}{2}R\right)
\]
with $R$ as the standard normal, denoted $\mathcal{N}\left(  0,1\right)  ,$
and expanding $G$ in a Taylor series formally leads to $\left(  \ref{CC}%
\right)  $ in the limit (see \cite{CC2} for details). One important
consequence of this equation is that the volatility (as defined in Section 5)
will be a function of $D/S,$ as is the price derivative. Prior research
\cite{CC2} concludes that volatility will be a minimum at price extrema, and a
maximum when the ratio $D/S$ (and consequently the magnitude of the price
derivative) are at a maximum. Of course, for market bottoms, there may be
other factors at work, e.g., margin calls, that would need to be modeled, as
discussed in Section 5. Equation $\left(  \ref{CC}\right)  $ has important
implications for risk assessment and options pricing. For example, if one is
using the classical equation $\left(  \ref{BS}\right)  ,$ and measuring
$\sigma,$ the risk and volatility would be underestimated if the trading price
of the asset is near a market top. There is empirical evidence that major
stock market tops are associated with low volatility (Sornette et. al., 2018
\cite{S}) suggestive of the maxim "calm before the storm." The supply,
$S\left(  t\right)  $ and demand, $D\left(  t\right)  $, in $\left(
\ref{CC}\right)  $ can be specified, or be coupled to other differential
equations, such as the asset flow equations that have been developed since the
late 1980's (see e.g., Caginalp and Balenovich, 1999 \cite{CB}, Merdan and
Alisen, 2011 \cite{MA}, DeSantis and Swigon, 2018 $\cite{DW},$ and references therein).

While volatility is an active research area (see e.g., \cite{FM}, \cite{HW},
\cite{SD}, \cite{SS}) and advances in empirical calculations of volatility
have been made in recent years, an approach based on microeconomics and
$\left(  \ref{CC}\right)  $ can be instrumental in an integrated understanding
of price change and volatility.

In this paper, we present a rigorous derivation of the probability density
corresponding to $\left(  \ref{CC}\right)  .$ This provides a justification of
the fat tail properties established in \cite{CC} that demonstrated that a
power law decay in relative price change is a consequence of the probability
distribution in inherent in the supply/demand ratio. In other words, assuming
that supply and demand are normally distributed (as one would expect from the
Central Limit Theorem) the quotient will have a density that decays (in many
cases) as a power law. Furthermore, we use this density to derive the
stochastic equation $\left(  \ref{CC}\right)  $.

\bigskip

\bigskip

\textbf{2. Derivation of the density for stochastic asset dynamics equation.
\ }In this section we derive rigorously the density of the relative price
change within $\left(  t,t+\Delta t\right)  $ that is generated by $\left(
\ref{CC}\right)  .$ The exact result is given by $\left(  \ref{f3}\right)  $ below.

First, we define the requirements for a function $G:\mathbb{R}^{+}%
\mathbb{\rightarrow}\mathbb{R}$ that will specify the nature of the relative
price change as a function of the ratio of demand to supply. In order to be
compatible with basic ideas of economics, $G$ must be increasing (see $\left(
ii\right)  $ below). For $S$ and $D$ to be on an equal footing, one needs the
condition $\left(  iii\right)  $ below. The other conditions have been
discussed in \cite{CC}.

\bigskip

\textbf{2.1 Condition} $\boldsymbol{G.}$ \ The function $G:\mathbb{R}%
^{+}\rightarrow\mathbb{R}$ is a twice continuously differentiable function satisfying

$\left(  i\right)  $ $G\left(  1\right)  =0,$\ $\left(  ii\right)  $
$G^{\prime}\left(  x\right)  >0$ \ all $x\in\mathbb{R}^{+},\ \left(
iii\right)  \ G\left(  x\right)  =-G\left(  \frac{1}{x}\right)  ,\ $%
\[
\left(  iv\right)  \ \ \lim_{x\rightarrow\infty}xG^{\prime}\left(  x\right)
=\infty\ \ \text{and}\ \ \ \lim_{x\rightarrow0+}xG^{\prime}\left(  x\right)
=\infty.
\]

\[
\left(  v\right)  \ \left(  xG^{\prime}\left(  x\right)  \right)  ^{\prime
}\ \ is\ \ \left\{
\begin{array}
[c]{ccc}%
<0 & if & x<1\\
>0 & if & x>1
\end{array}
\right.  .\ \ ///
\]
These properties clearly imply the following two relations:%
\begin{equation}
xG^{\prime}\left(  x\right)  =\frac{1}{x}G^{\prime}\left(  \frac{1}{x}\right)
. \label{deriv}%
\end{equation}%
\begin{equation}
\lim_{x\rightarrow\infty}G\left(  x\right)  =\infty. \label{inf}%
\end{equation}

In deriving the stochastic equations for price change, additional conditions
are needed, and listed in the augmented condition below.

\bigskip

\textbf{Condition} $\boldsymbol{G}_{A}\boldsymbol{.}$ Let $G$ satisfy
Condition $G,$ and, in addition, assume that $G^{-1}$ has four continuous
derivatives that are bounded on bounded subsets of the domain.

\bigskip

\textbf{2.2.} \textbf{Examples of functions that satisfy Condition }$G_{A}%
$\textbf{.}

One can readily verify that the following functions satisfy this condition:

$\left(  i\right)  $ $G\left(  x\right)  =x^{q}-x^{-q}$ for $q>0;$

$\left(  ii\right)  $ $G\left(  x\right)  =\left(  x-x^{-1}\right)  ^{q}$ for
$q$ an odd positive integer.

Note that when $x-1=\frac{D-S}{S}$ is small, the basic model $\left(
i\right)  $ with $q:=1$ is the continuous analog of the simple excess demand
model often considered in economics, i.e., $g\left(  \frac{D}{S}\right)
=\frac{D-S}{S},$ in $\left(  \ref{nonlin discrete p}\right)  $.

\bigskip

\textbf{2.3.} \textbf{The derivation.} We denote by $\tilde{D}$ the total
demand including randomness per unit time. During a time interval $\left(
t,t+\Delta t\right)  $, the total demand is $\tilde{D}\Delta t$ while the
deterministic component is $D\Delta t$, where $D$ (which can depend on time,
price and other factors) can be regarded as the expected value of $\tilde{D}$.
In the same way, the supply $\tilde{S}$ and expectation of supply, $S$ are defined.

We can then write the total demand during time $\Delta t$ divided by the total
supply in that time period as%
\[
\frac{\tilde{D}\Delta t}{\tilde{S}\Delta t}.
\]
We argue below that given a large number of decision makers, the Central Limit
Theorem states, under broad conditions, that the randomness in the buy orders
placed by many different agents will be approximately Gaussian (as discussed
further in Section 2.3). Note that this is a very different -- and more easily
justified -- assumption than the hypothesis that stock prices changes are
log-normal. There is no convincing argument that asset price changes should
satisfy the hypotheses of the Central Limit Theorem, since price adjustments
evolve in a complex manner through the supply and demand.

\bigskip

This does not mean that the supply and demand will be centered around values
that would be consistent with fundamental value. For example, if there is a
very positive image portrayed about some company, we still expect a normal
distribution in the demand as a function of the price that the potential
buyers are willing to pay, though that price may be much higher than would be
indicated by finance value calculations such as the potential dividend stream.
Also, there are, of course, correlations among groups of investors. However,
for actively traded stocks, there are such a large number of such groups (some
focusing on trend, others on fundamentals, etc.) which act independently that
these correlations would cease to be relevant. Casual observations indicate
that supply and demand, like may other preference issues, have a distribution
that is qualitatively similar to a Gaussian, except near the tail, which is
usually irrelevant since it is far from the crossing of supply and demand.
While many studies have been performed on the distribution of relative price
changes, more study is needed for the distribution of supply and demand
distribution. Tests on their normality are the ultimate criterion for
determining the validity of the assumption on an empirical basis.

We consider first the impact of randomness on supply and demand. As discussed
in Appendix A, it is the "market orders" rather than the "limit orders" that
are mainly responsible for changes in price. Since there are many independent
agents using publicly available information to make decisions on trading a
particular asset, we can assume that the orders are independent and
identically distributed with a given mean and variance. During this time
interval there will be additional demand (positive or negative) due to
randomness from a variety of sources such as news items that alter the value
or desirability of the asset. For example, if there is an unexpected
announcement that the Federal Reserve is lowering interest rates, it may
increase the demand for a stock index by $1\%.$ In other words, the random
term in the demand is proportional to the baseline deterministic demand (i.e.,
the expected demand during this time interval, $D\Delta t$). Denoting by
$\frac{\sigma}{2}R$ this random term per unit demand and per unit time, one
has then that the total random factor in demand is given by $\frac{\sigma}%
{2}DR\Delta t$ and the total demand, $\tilde{D}\Delta t$, and supply,
$\tilde{S}\Delta t$, during the time interval $\left(  t,t+\Delta t\right)  $
is given by%
\begin{equation}
\tilde{D}\Delta t=D\Delta t\left(  1+\frac{\sigma}{2}R\right)  ,\ \ \ \tilde
{S}\Delta t=S\Delta t\left(  1-\frac{\sigma}{2}R\right)  .
\label{demand/supply}%
\end{equation}
The simple assumption is that a random event that increases demand for the
asset decreases supply. This assumption can be relaxed as discussed below so
that any correlation between the two random variables\ (for supply and demand,
respectively) can be considered.

The random term, $\frac{\sigma}{2}R\Delta t$, is actually the average of a
large number of agents with a distribution that we cannot necessarily specify.
However, the Central Limit Theorem indicates that the limiting distribution
will be normal since a sufficiently large number of them, having the same
public information, can be assumed to have a particular distribution. One can
then expect that $R$ will be a standard normal, often expressed as
$\mathcal{N}\left(  0,1\right)  ,$ denoting a Gaussian with mean $0$ and
variance $1,$ so that the term $D\Delta t\frac{\sigma}{2}R$ is normal with
mean zero and variance $\left(  D\Delta t\frac{\sigma}{2}\right)  ^{2}.$

If we accept the standard assumptions (see Schuss, 2009 \cite{SC}, p. 39) of
the independence of the random events on $\left(  t,t+\Delta t\right)  $ from
events prior to $t,$ as well as the appropriate continuity properties in the
limit as $\Delta t\rightarrow0,$ then we can assume that $R\Delta t=\Delta
W:=W\left(  t+\Delta t\right)  -W\left(  t\right)  .$ Summarizing, we write
the total demand and supply on this interval as%
\begin{equation}
\tilde{D}\Delta t=D\Delta t+\frac{\sigma}{2}D\Delta W,\ \ \ \tilde{S}\Delta
t=S\Delta t-\frac{\sigma}{2}S\Delta W. \label{total}%
\end{equation}
Given, as noted above, that $R$ can be interpreted as the fractional random
component per unit time. This is consistent with the interpretation of $R$ as
$\Delta W/\Delta t$ which, in the limit as $\Delta t\rightarrow0,$ is "white
noise"\ or the derivative of Brownian motion, which does not exist in the
classical mathematical sense. Nevertheless, we do not need to utilize the
limit of $\Delta W/\Delta t$ concept here.

The relative change in price $P^{-1}dP/dt,$ in the time interval $\Delta t,$
is then postulated to be proportional to a function, $G$ of the total demand
divided by supply during this time interval, i.e.,
\begin{equation}
P^{-1}\frac{\Delta P}{\Delta t}=G\left(  \frac{D\Delta t\left(  1+\frac
{\sigma}{2}R\right)  }{S\Delta t\left(  1-\frac{\sigma}{2}R\right)  }\right)
. \label{nonlin general}%
\end{equation}
We assume that $G$ satisfies condition $G_{A}$ and incorporates the time constant.

Using the standard normal random variable, $Y\sim\mathcal{N}\left(
0,1\right)  $ this equation has the form%

\begin{align}
P^{-1}\frac{\Delta P}{\Delta t}  &  =G\left(  \frac{D}{S}\frac{\Delta
t+\frac{\sigma}{2}Y\left(  \Delta t\right)  ^{1/2}}{\Delta t-\frac{\sigma}%
{2}Y\left(  \Delta t\right)  ^{1/2}}\right) \label{nonlin cont}\\
&  \sim G\left(  \frac{\mathcal{N}\left(  D\Delta t,D^{2}\frac{\sigma^{2}}%
{4}\Delta t\right)  }{\mathcal{N}\left(  S\Delta t,S^{2}\frac{\sigma^{2}}%
{4}\Delta t\right)  }\right) \nonumber
\end{align}
with numerator and denominator anti-correlated. The more general assumption
that the supply and demand have a correlation different from $-1$ can be
considered by assuming
\begin{equation}
P^{-1}\frac{\Delta P}{\Delta t}=G\left(  \frac{D}{S}\frac{\Delta
t+\frac{\sigma}{2}Y_{1}\left(  \Delta t\right)  ^{1/2}}{\Delta t+\frac{\sigma
}{2}Y_{2}\left(  \Delta t\right)  ^{1/2}}\right)  \label{general correlation}%
\end{equation}
where $Y_{1}$ and $Y_{2}$ are both standard normals, i.e., $\mathcal{N}\left(
0,1\right)  $, but have some correlation $\rho>-1.$ The analysis is then more
complicated, but the essential ideas are similar.

We proceed with the analysis of the basic equation $\left(  \ref{nonlin cont}%
\right)  $ by defining random variables in order to obtain the density
function for the random variable on the right hand side.

Toward this end, define%
\begin{align*}
X  &  :=\frac{\Delta t+\frac{\sigma}{2}Y\left(  \Delta t\right)  ^{1/2}%
}{\Delta t-\frac{\sigma}{2}Y\left(  \Delta t\right)  ^{1/2}},\ \ \ X_{1}%
:=\frac{D}{S}\frac{\Delta t+\frac{\sigma}{2}Y\left(  \Delta t\right)  ^{1/2}%
}{\Delta t-\frac{\sigma}{2}Y\left(  \Delta t\right)  ^{1/2}}\\
X_{2}  &  :=G\left(  \frac{D}{S}\frac{\Delta t+\frac{\sigma}{2}Y\left(  \Delta
t\right)  ^{1/2}}{\Delta t-\frac{\sigma}{2}Y\left(  \Delta t\right)  ^{1/2}%
}\right)  ,\ \ X_{3}:=G\left(  \frac{D}{S}\frac{\Delta t+\frac{\sigma}%
{2}Y\left(  \Delta t\right)  ^{1/2}}{\Delta t-\frac{\sigma}{2}Y\left(  \Delta
t\right)  ^{1/2}}\right)  \Delta t
\end{align*}%
\[
X_{1}=\frac{D}{S}X,\ \ X_{2}=G\left(  X_{1}\right)  ,\ \ X_{3}=G\left(
X_{1}\right)  \Delta t.
\]
The price equation can then be written as
\begin{equation}
P^{-1}\Delta P=G\left(  X_{1}\right)  \Delta t=X_{3}. \label{price disc}%
\end{equation}

This means that the change in the relative price is governed by the random
variable $X_{3},$ whose density we compute below, and show that it is
approximately normal\footnote{See Tong, 1990 \cite{TO} for a comprehensive
exposition of the properties of the normal distribution} for small $\sigma$
and fixed $\Delta t.$ The first of the random variables above, namely $X$, is
a quotient of two anticorrelated normal random variables for which one has an
exact expression. In particular, Theorem 4.3 of Caginalp and Caginalp, 2018
\cite{CC} can be applied\ to $X$ in the form
\[
X=\frac{1+\frac{\sigma}{2\left(  \Delta t\right)  ^{1/2}}Y}{1-\frac{\sigma
}{2\left(  \Delta t\right)  ^{1/2}}Y}.
\]
to yield the density of $X$ as%
\[
f_{X}\left(  x\right)  =\frac{\frac{\sigma}{\left(  \Delta t\right)  ^{1/2}}%
}{\sqrt{2\pi}}\frac{e^{-\frac{1}{2}\frac{\left(  x-1\right)  ^{2}}{\frac
{1}{\Delta t}\left(  \frac{\sigma}{2}\right)  ^{2}\left(  x+1\right)  ^{2}}}%
}{\frac{1}{\left(  \Delta t\right)  }\left(  \frac{\sigma}{2}\right)
^{2}\left(  x+1\right)  ^{2}}.
\]

The density of $X_{1}:=\frac{D}{S}X$ is then calculated as%

\begin{align}
f_{X_{1}}\left(  x\right)   &  =\frac{f_{X}\left(  \frac{x}{D/S}\right)
}{D/S}=\frac{\frac{\sigma}{\left(  \Delta t\right)  ^{1/2}}}{\sqrt{2\pi}%
\frac{D}{S}}\frac{e^{-\frac{1}{2}\frac{\left(  \frac{x}{D/S}-1\right)  ^{2}%
}{\frac{1}{\Delta t}\left(  \frac{\sigma}{2}\right)  ^{2}\left(  \frac{x}%
{D/S}+1\right)  ^{2}}}}{\frac{1}{\Delta t}\left(  \frac{\sigma}{2}\right)
^{2}\left(  \frac{x}{D/S}+1\right)  ^{2}}\nonumber\\
&  =\frac{1}{\sqrt{2\pi}\frac{D}{S}\frac{\sigma}{\left(  \Delta t\right)
^{1/2}}}\frac{e^{-\frac{1}{2}\frac{\left(  \frac{x}{D/S}-1\right)  ^{2}}%
{\frac{\sigma^{2}}{4\left(  \Delta t\right)  }\left(  \frac{x}{D/S}+1\right)
^{2}}}}{\frac{1}{4}\left(  \frac{x}{D/S}+1\right)  ^{2}}. \label{III.1}%
\end{align}
Note that when $x\approx D/S$ the factor $\left(  \frac{x}{D/S}+1\right)
^{2}\approx4$ cancels the $1/4,$ so that $f_{X_{1}}\left(  x\right)  $ is
approximately normal. However, we continue with the exact expression above,
which can be expressed as%
\begin{equation}
f_{X_{1}}\left(  x\right)  =\frac{1}{\sqrt{2\pi}\frac{D}{S}\frac{\sigma
}{\left(  \Delta t\right)  ^{1/2}}}\frac{e^{-\frac{1}{2}\frac{\left(
x-D/S\right)  ^{2}}{\frac{\sigma^{2}}{4\left(  \Delta t\right)  }\left(
x+D/S\right)  ^{2}}}}{\frac{1}{4}\left(  \frac{S}{D}\right)  ^{2}\left(
x+D/S\right)  ^{2}}. \label{III.2}%
\end{equation}
For $G:\mathbb{R}^{+}\mathbb{\rightarrow}\mathbb{R}$ satisfying Condition
$G_{A}$, the density, $f_{2}$, of $X_{2}:=G\left(  X_{1}\right)  $ is given
by
\[
f_{2}\left(  y\right)  =\frac{f_{1}\left(  G^{-1}\left(  y\right)  \right)
}{G^{\prime}\left(  G^{-1}\left(  y\right)  \right)  }=\frac{f_{1}\left(
x\right)  }{G^{\prime}\left(  x\right)  }%
\]
where we use the notation $y=G\left(  x\right)  $, $G^{-1}\left(  y\right)
=x.$ Substitution into $\left(  \ref{III.2}\right)  $ yields the exact
expression%
\begin{equation}
f_{2}\left(  y\right)  =\frac{1}{\sqrt{2\pi}\frac{S}{D}\frac{\sigma}{\left(
\Delta t\right)  ^{1/2}}G^{\prime}\left(  G^{-1}\left(  y\right)  \right)
}\frac{e^{-\frac{1}{2}\frac{\left(  G^{-1}\left(  y\right)  -D/S\right)  ^{2}%
}{\frac{\sigma^{2}}{4\left(  \Delta t\right)  }\left(  G^{-1}\left(  y\right)
+D/S\right)  ^{2}}}}{\frac{1}{4}\left(  G^{-1}\left(  y\right)  +D/S\right)
^{2}}. \label{III.3}%
\end{equation}
Finally, the density $f_{3}$ of $X_{3}:=X_{2}\Delta t$ , the quantity
governing price change in $\left(  \ref{price disc}\right)  $ is given by
$f_{3}\left(  y\right)  =f_{2}\left(  \frac{y}{\Delta t}\right)  /\Delta t,$
\ so substitution into $\left(  \ref{III.3}\right)  $ yields the exact density
given by the following.

\bigskip

\textbf{Proposition 2.1.} Let $G:\mathbb{R}^{+}\rightarrow\mathbb{R}$ be a
function satisfying Condition $G_{A}$ and $Y$ \ be defined as the standard
normal, $\mathcal{N}\left(  0,1\right)  $. Then the density of the random
variable that constitutes the right hand side of the price equation $\left(
\ref{price disc}\right)  $%
\begin{equation}
G\left(  \frac{D}{S}\frac{1+\frac{\sigma}{2\left(  \Delta t\right)  ^{1/2}}%
Y}{1-\frac{\sigma}{2\left(  \Delta t\right)  ^{1/2}}Y}\right)  \Delta t,
\label{G}%
\end{equation}
is given by
\begin{equation}
f_{3}\left(  y\right)  =\frac{1}{\sqrt{2\pi}\frac{S}{D}\frac{\sigma\Delta
t}{\left(  \Delta t\right)  ^{1/2}}G^{\prime}\left(  G^{-1}\left(  \frac
{y}{\Delta t}\right)  \right)  }\frac{e^{-\frac{1}{2}\frac{\left(
G^{-1}\left(  \frac{y}{\Delta t}\right)  -D/S\right)  ^{2}}{\frac{\sigma^{2}%
}{4\left(  \Delta t\right)  }\left(  G^{-1}\left(  \frac{y}{\Delta t}\right)
+D/S\right)  ^{2}}}}{\frac{1}{4}\left(  G^{-1}\left(  \frac{y}{\Delta
t}\right)  +D/S\right)  ^{2}}. \label{f3}%
\end{equation}
\bigskip

\bigskip

\bigskip

\textbf{3. Asymptotics of the density. }The density $\left(  \ref{f3}\right)
$ can be studied asymptotically for small variance, i.e., $\sigma^{2}<<1.$ In
this section, we obtain an exact expression that will be used subsequently for
asymptotic analysis. Toward this end, we express the terms involving $G^{-1}$
in an exact Maclaurin-Taylor expansion with remainder with $\zeta_{1}$ and
$\zeta_{2}$ as the usual intermediate values between $y/\Delta t-y_{0}/\Delta
t$ and define $\eta_{1}$ by $G\left(  \eta_{1}\right)  =\zeta_{1}:$
\begin{align}
G^{-1}\left(  \frac{y}{\Delta t}\right)  -\frac{D}{S}  &  =G^{-1}\left(
\frac{y}{\Delta t}\right)  -G^{-1}\left(  \frac{y_{0}}{\Delta t}\right)
\label{Ginv expansion}\\
&  =\left(  G^{-1}\right)  ^{\prime}\left(  \zeta_{1}\right)  \left(  \frac
{y}{\Delta t}-\frac{y_{0}}{\Delta t}\right) \nonumber\\
&  =\frac{1}{G^{\prime}\left(  \eta_{1}\right)  }\left(  y-G\left(
D/S\right)  \Delta t\right)  \frac{1}{\Delta t}\nonumber
\end{align}
so that one has also%
\begin{equation}
G^{-1}\left(  \frac{y}{\Delta t}\right)  +\frac{D}{S}=2\frac{D}{S}+\frac
{1}{G^{\prime}\left(  \eta_{1}\right)  }\left(  y-G\left(  D/S\right)  \Delta
t\right)  \frac{1}{\Delta t}. \label{Ginv expansion2}%
\end{equation}

For the numerator of the exponent we need the next term in the series
expansion:%
\begin{align*}
G^{-1}\left(  \frac{y}{\Delta t}\right)   &  =G^{-1}\left(  \frac{y_{0}%
}{\Delta t}\right)  +\left(  G^{-1}\right)  ^{\prime}\left(  \frac{y_{0}%
}{\Delta t}\right)  \left(  \frac{y}{\Delta t}-\frac{y_{0}}{\Delta t}\right)
\\
&  +\frac{\left(  G^{-1}\right)  ^{\prime\prime}\left(  \zeta_{2}\right)  }%
{2}\left(  \frac{y}{\Delta t}-\frac{y_{0}}{\Delta t}\right)  ^{2}%
\end{align*}
and rewriting, we have%
\begin{align}
G^{-1}\left(  \frac{y}{\Delta t}\right)   &  =\frac{D}{S}+\frac{1}{G^{\prime
}\left(  \frac{D}{S}\right)  }\left(  y-G\left(  D/S\right)  \Delta t\right)
\frac{1}{\Delta t}\nonumber\\
&  +\frac{\left(  G^{-1}\right)  ^{\prime\prime}\left(  \zeta_{2}\right)  }%
{2}\frac{\left(  y-G\left(  D/S\right)  \Delta t\right)  ^{2}}{\left(  \Delta
t\right)  ^{2}}. \label{Ginv expansion3}%
\end{align}
From Condition $G_{A}$, the first and second derivatives $G^{-1}$ are bounded.
Writing
\[
f_{3}=:e^{E}/B
\]%
\[
E:=-\frac{1}{2}\frac{\left(  G^{-1}\left(  \frac{y}{\Delta t}\right)
-G^{-1}\left(  \frac{y_{0}}{\Delta t}\right)  \right)  ^{2}}{\frac{\sigma^{2}%
}{\Delta t}\frac{1}{4}\left(  G^{-1}\left(  \frac{y}{\Delta t}\right)
+G^{-1}\left(  \frac{y_{0}}{\Delta t}\right)  \right)  ^{2}}%
\]
we examine the terms $E$ and $B$ below, so that substitution of the expansions
$\left(  \ref{Ginv expansion}\right)  ,$ $\left(  \ref{Ginv expansion2}%
\right)  $ in numerator and denominator yields%
\[
E=-\frac{1}{2}\frac{\left(  \frac{1}{G^{\prime}\left(  \frac{D}{S}\right)
}\left(  y-G\left(  D/S\right)  \Delta t\right)  \frac{1}{\Delta t}%
+\frac{\left(  G^{-1}\right)  ^{\prime\prime}\left(  \zeta_{2}\right)  }%
{2}\frac{\left(  y-G\left(  D/S\right)  \Delta t\right)  ^{2}}{\left(  \Delta
t\right)  ^{2}}\right)  ^{2}}{\frac{\sigma^{2}}{\Delta t}\frac{1}{4}\left(
2\frac{D}{S}+\frac{1}{G^{\prime}\left(  \eta_{1}\right)  }\left(  y-G\left(
D/S\right)  \Delta t\right)  \frac{1}{\Delta t}\right)  ^{2}}.
\]
Factoring the key terms, one has
\begin{equation}
E=-\frac{1}{2}\frac{\left(  \left(  y-G\left(  D/S\right)  \Delta t\right)
+\frac{G^{\prime}\left(  \frac{D}{S}\right)  \left(  G^{-1}\right)
^{\prime\prime}\left(  \zeta_{2}\right)  }{2}\frac{\left(  y-G\left(
D/S\right)  \Delta t\right)  ^{2}}{\left(  \Delta t\right)  }\right)  ^{2}%
}{\left[  G^{\prime}\left(  \frac{D}{S}\right)  \right]  ^{2}\sigma^{2}\Delta
t\left(  \frac{D}{S}\right)  ^{2}\left(  1+\frac{1}{2G^{\prime}\left(
\eta_{1}\right)  D/S}\left(  y-G\left(  D/S\right)  \Delta t\right)  \frac
{1}{\Delta t}\right)  ^{2}}. \label{E}%
\end{equation}

We will analyze this expression in the next section to show that (for fixed
$\Delta t$) the terms beyond leading order can be controlled. Next we
manipulate the expression for $B$, the denominator of $f_{3}$ given by
$\left(  \ref{f3}\right)  :$%
\begin{align}
B  &  :=\sqrt{2\pi}\frac{S}{D}\frac{\sigma\Delta t}{\left(  \Delta t\right)
^{1/2}}G^{\prime}\left(  G^{-1}\left(  \frac{y}{\Delta t}\right)  \right)
\frac{1}{4}\left(  G^{-1}\left(  \frac{y}{\Delta t}\right)  +D/S\right)
^{2}\nonumber\\
&  =\sqrt{2\pi}\frac{S}{D}\sigma\left(  \Delta t\right)  ^{1/2}G^{\prime
}\left(  G^{-1}\left(  \frac{y}{\Delta t}\right)  \right)  \left(  \frac{D}%
{S}\right)  ^{2}\cdot\label{Bo}\\
&  \left(  1+\frac{1}{2\frac{D}{S}G^{\prime}\left(  \eta_{1}\right)  }\left(
y-G\left(  D/S\right)  \Delta t\right)  \frac{1}{\Delta t}\right)
^{2}.\nonumber
\end{align}
Using the series expansion for $G^{-1}\left(  \frac{y}{\Delta t}\right)  $,
namely, $\left(  \ref{Ginv expansion}\right)  $ together with the expansion
for $G^{\prime}$ we can write the Maclaurin-Taylor expression,
\begin{align*}
G^{\prime}\left(  G^{-1}\left(  \frac{y}{\Delta t}\right)  \right)   &
=G^{\prime}\left(  \frac{D}{S}+\frac{1}{G^{\prime}\left(  \eta_{1}\right)
}\left(  y-G\left(  D/S\right)  \Delta t\right)  \frac{1}{\Delta t}\right) \\
&  =G^{\prime}\left(  \frac{D}{S}\right)  +G^{\prime\prime}\left(  \xi\right)
\left(  \frac{1}{G^{\prime}\left(  \eta_{1}\right)  }\left(  y-G\left(
D/S\right)  \Delta t\right)  \frac{1}{\Delta t}\right)
\end{align*}
in terms of another intermediate value $\xi$ between $y/\Delta t$ and
$y_{0}/\Delta t.$ Substitution into $\left(  \ref{Bo}\right)  $ and some
algebraic simplification yields%
\begin{align}
B  &  =\sqrt{2\pi}\frac{D}{S}\sigma\left(  \Delta t\right)  ^{1/2}\left\{
G^{\prime}\left(  \frac{D}{S}\right)  +G^{\prime\prime}\left(  \xi\right)
\left(  \frac{1}{G^{\prime}\left(  \eta_{1}\right)  }\left(  y-G\left(
D/S\right)  \Delta t\right)  \frac{1}{\Delta t}\right)  \right\} \nonumber\\
&  \cdot\left(  1+\frac{1}{2\frac{D}{S}G^{\prime}\left(  \eta_{1}\right)
}\left(  y-G\left(  D/S\right)  \Delta t\right)  \frac{1}{\Delta t}\right)
^{2}. \label{B}%
\end{align}

Having used the Taylor expansions to rewrite $E$ and $B,$ we have an exact
form of $f_{3}=:e^{E}/B$ that is the density for the variable expressed in
$\left(  \ref{G}\right)  $. We will analyze these terms further in the next section.

\bigskip

\bigskip

\bigskip

\textbf{4. Analysis of the limits. }Our next objective is to analyze the
density, $f_{3},$ obtained above for the random variable, $G\left(
...\right)  \Delta t$, in $\left(  \ref{nonlin cont}\right)  ,$ that is an
expression of the density of relative price changes, through $\left(
\ref{nonlin general}\right)  ,$ in a time interval $\left(  t,t+\Delta
t\right)  .$

One implication of this analysis will be to establish rigorously the power-law
tail of the density of relative price change. We will also compare the density
$f_{3}$ for $X_{3}$ (given by $\left(  \ref{f3}\right)  $) with the normal
approximation denoted by $X_{3}^{\left(  N\right)  }$ and defined below in
$\left(  \ref{normal}\right)  .$ One feature of the quotient of normals that
has long been established is that its density near the mean is close to a
normal under a broad set of assumptions on the parameters. We will see in
Section 5 that an implication of this idea can be used to derive stochastic
differential equations for relative price change.

A tool we will use below is the basic result in asymptotic Laplace integrals
for a continuous, bounded function $S$ and a twice continuously differentiable
function $h$ with maximum at $0$. Assuming $a>>1$, one has
\begin{equation}
I:=\int_{-\infty}^{\infty}u\left(  x\right)  e^{ah\left(  x\right)
}dx=u\left(  0\right)  \left(  \frac{-2\pi}{ah^{\prime\prime}\left(  0\right)
}\right)  ^{1/2}e^{ah\left(  0\right)  }+e^{ah\left(  0\right)  }O\left(
a^{-3/2}\right)  \label{a}%
\end{equation}
(see e.g., Murray, 2012 \cite{M} p. 34, and the classic text by De Bruijn,
1981 \cite{D}). Our goal is to prove that the expectation of an arbitrary
bounded, continuous function $R$ with respect to $X_{3}$ is given by the
expectation of $R$ with respect to
\begin{equation}
X_{3}^{\left(  N\right)  }\sim\mathcal{N}\left(  G\left(  D/S\right)  \Delta
t,\sigma^{2}\Delta t\left(  G^{\prime}\left(  \frac{D}{S}\right)  \frac{D}%
{S}\right)  ^{2}\right)  \label{normal}%
\end{equation}
plus terms of order $\sigma^{2}$.

\bigskip

\textbf{Theorem 4.1.} Let $f_{3}$ and $f_{3}^{\left(  N\right)  }$ be defined
as the densities of $X_{3}$ and $X_{3}^{\left(  N\right)  },$ defined by
$\left(  \ref{f3}\right)  $ and $\left(  \ref{normal}\right)  .$ For fixed
$\Delta t$ and an arbitrary bounded, continuous function $R,$ one has%
\begin{equation}
\left\vert \int_{-\infty}^{\infty}R\left(  x\right)  f_{3}\left(  x\right)
dx-\int_{-\infty}^{\infty}R\left(  x\right)  f_{3}^{\left(  N\right)  }\left(
x\right)  dx\right\vert \leq C\sigma^{2} \label{conv}%
\end{equation}
where $C$ is a constant that is independent of $\sigma^{2}.$

\bigskip

\textbf{Remark.} Inequality $\left(  \ref{conv}\right)  $ implies that $X_{3}$
converges as $\sigma\rightarrow0$ to $X_{3}^{\left(  N\right)  }$ in
distribution (see Appendix D).

\bigskip

Proof. $\ \left(  i\right)  $ First we consider the limiting integral%

\[
I_{1}=\int_{-\infty}^{\infty}R\left(  y\right)  f_{3}^{\left(  N\right)
}\left(  y\right)  dy.
\]
We have then, with $z:=y-G\left(  D/S\right)  \Delta t$ and $\tilde{R}\left(
z\right)  =\tilde{R}\left(  y-G\left(  D/S\right)  \Delta t\right)  =R\left(
y\right)  ,$ the expressions%
\begin{align*}
I_{1}  &  =\frac{1}{\sigma\left(  \Delta t\right)  ^{1/2}}\int_{-\infty
}^{\infty}R\left(  y\right)  \frac{e^{-\frac{1}{2}\frac{\left(  y-G\left(
D/S\right)  \Delta t\right)  ^{2}}{\sigma^{2}\Delta t\left(  G^{\prime}\left(
\frac{D}{S}\right)  \frac{D}{S}\right)  ^{2}}}}{\sqrt{2\pi}G^{\prime}\left(
\frac{D}{S}\right)  \frac{D}{S}}dy\\
&  =\frac{1}{\sigma\left(  \Delta t\right)  ^{1/2}}\int_{-\infty}^{\infty
}\tilde{R}\left(  z\right)  \frac{e^{-\frac{1}{2}\frac{z^{2}}{\sigma^{2}\Delta
t\left(  G^{\prime}\left(  \frac{D}{S}\right)  \frac{D}{S}\right)  ^{2}}}%
}{\sqrt{2\pi}G^{\prime}\left(  \frac{D}{S}\right)  \frac{D}{S}}dz.
\end{align*}
Let $a:=\frac{1}{\sigma^{2}\Delta t}$ so $a>>1,$ and%
\[
u\left(  z\right)  :=\frac{\tilde{R}\left(  z\right)  }{\sqrt{2\pi}G^{\prime
}\left(  \frac{D}{S}\right)  \frac{D}{S}}%
\]%
\[
h\left(  z\right)  :=-\frac{1}{2}\frac{z^{2}}{\left(  G^{\prime}\left(
\frac{D}{S}\right)  \frac{D}{S}\right)  ^{2}},
\]
so differentiating twice leads to%
\[
h^{\prime\prime}\left(  0\right)  =-\frac{1}{\left(  G^{\prime}\left(
\frac{D}{S}\right)  \frac{D}{S}\right)  ^{2}}.
\]
Substitution into $\left(  \ref{a}\right)  $ yields%
\begin{align*}
I_{1}  &  =a^{1/2}u\left(  0\right)  \left(  \frac{-2\pi}{ah^{\prime\prime
}\left(  0\right)  }\right)  ^{1/2}+O\left(  a^{-3/2}\right) \\
&  =\tilde{R}\left(  0\right)  \frac{1}{\sqrt{2\pi}G^{\prime}\left(  \frac
{D}{S}\right)  \frac{D}{S}}\sqrt{2\pi}\left(  G^{\prime}\left(  \frac{D}%
{S}\right)  \frac{D}{S}\right)  ^{-1}+O\left(  \sigma^{2}\Delta t\right) \\
&  =R\left(  D/S\right)  +O\left(  \sigma^{2}\Delta t\right)  .
\end{align*}

$\bigskip$

$\left(  ii\right)  $ Next, using the notation $\hat{R}\left(  z\right)
:=R\left(  y\right)  ,$\ consider the integral
\begin{align*}
I_{2}  &  =\int_{-\infty}^{\infty}R\left(  y\right)  f_{3}\left(  y\right)
dy\\
&  =\int_{-\infty}^{\infty}\hat{R}\left(  z\right)  \frac{e^{ah_{1}\left(
z\right)  }}{B_{1}}dz
\end{align*}
where $h_{1}$, $B_{1}$ and $u_{1}$ are defined by
\begin{equation}
h_{1}\left(  z\right)  :=-\frac{1}{2}\frac{\left(  z+\frac{G^{\prime}\left(
\frac{D}{S}\right)  \left(  G^{-1}\right)  ^{\prime\prime}\left(  \zeta
_{2}\right)  }{2}\frac{z^{2}}{\left(  \Delta t\right)  }\right)  ^{2}}{\left[
G^{\prime}\left(  \frac{D}{S}\right)  \frac{D}{S}\right]  ^{2}\left(
1+\frac{1}{2G^{\prime}\left(  \eta_{1}\right)  D/S}z\frac{1}{\Delta t}\right)
^{2}} \label{h1}%
\end{equation}%
\[
B_{1}\left(  z\right)  :=\sqrt{2\pi}G^{\prime}\left(  G^{-1}\left(  \frac
{y}{\Delta t}\right)  \right)  \frac{D}{S}\left(  1+\frac{1}{2\frac{D}%
{S}G^{\prime}\left(  \eta_{1}\right)  }\frac{z}{\Delta t}\right)  ^{2}
\]%
\[
u_{1}\left(  z\right)  :=\frac{\hat{R}\left(  z\right)  }{B_{1}\left(
z\right)  }.
\]
Note that $B_{1}\left(  0\right)  =\sqrt{2\pi}\frac{D}{S}G^{\prime}\left(
\frac{D}{S}\right)  $ and $h_{1}^{\prime\prime}\left(  0\right)  =-\frac
{1}{\left[  \frac{D}{S}G^{\prime}\left(  \frac{D}{S}\right)  \right]  ^{2}}$
(see Appendix B). Note also that $\Delta t$ is a fixed constant. One has then%
\begin{align*}
I_{2}  &  =a^{1/2}g\left(  0\right)  \left(  \frac{-2\pi}{ah_{1}^{\prime
\prime}\left(  0\right)  }\right)  ^{1/2}+O\left(  a^{-3/2}\right) \\
&  =R\left(  D/S\right)  +O\left(  \sigma^{2}\Delta t\right)  .
\end{align*}

Since $I_{1}$ and $I_{2}$ differ by no more than a constant times $\sigma^{2}$
the inequality $\left(  \ref{conv}\right)  $ is thus established, so
expectations with respect to $X_{3}$ and $X_{3}^{\left(  N\right)  }$ are
within $O\left(  \sigma^{2}\right)  .$ $///$

\bigskip

\textbf{Remark.} This result tells us that with $\Delta t>0$ fixed, the
appropriately scaled normal density is an approximation to the density of
$G\Delta t$. This is despite the fact that the densities are quite different
at large values of the argument (i.e., exponential versus power law, or fat
tail, decay). As an alternative, one can choose the small parameter as
$\sigma\left(  \Delta t\right)  ^{-1/2}$ and allow $\Delta t$ to approach zero
as well, so long as $\sigma\left(  \Delta t\right)  ^{-1/2}\rightarrow0.$ This
is discussed in Appendix C.\bigskip

\bigskip

\textbf{5. The stochastic dynamics equation through the supply/demand
equation.}

\bigskip

\textbf{5.1. Volatility.} In this section we utilize the rigorous results
obtained thus far to write the expression for the stochastic process for
relative price change as well as expressions for volatility. Summarizing the
results of the previous section, we have%
\[
P^{-1}\frac{\Delta P}{\Delta t}=X_{2}\text{, \ i.e.,\ \ }P^{-1}\Delta P=X_{3}%
\]%
\[
X_{3}=X_{3}^{\left(  N\right)  }+O\left(  \sigma^{2}\right)
\]
where $X_{3}^{\left(  N\right)  }$ is normally distributed [see $\left(
\ref{normal}\right)  $ for definition]. Thus, the relative price change within
the time interval $\left(  t,t+\Delta t\right)  $ is close to a normal for
small variance,~$\sigma^{2}$, and attains the normal limit (in the sense of
distributions) as $\sigma\rightarrow0.$ Together these yield the expression
(to leading order in $\sigma$)%
\[
P^{-1}\Delta P\sim\mathcal{N}\left[  G\left(  \frac{D}{S}\right)  \Delta
t,\sigma^{2}\left(  \Delta t\right)  \left\{  \frac{D}{S}G^{\prime}\left(
\frac{D}{S}\right)  \right\}  ^{2}\right]  .
\]
which can also be expressed as%
\begin{equation}
P^{-1}\Delta P=G\left(  D/S\right)  \Delta t+\sigma G^{\prime}\left(
D/S\right)  \left(  D/S\right)  \Delta W, \label{discP}%
\end{equation}
where $\Delta W$ is normal with mean zero and variance $\Delta t.$

Using the basic theory (see e.g., Schuss, 2009 \cite{SC}) we set $\left[
W\left(  t\right)  ;t\geq0\right]  $ as the stochastic process defined on the
probability space $\left(  \Omega,\mathcal{F},\mathbb{P}\right)  $ that
satisfies the conditions of Brownian motion. In particular, $\mathbb{P}%
\left\{  W\left(  0\right)  =0\right\}  =1,$ the increments $\Delta
W:=W\left(  t+\Delta t\right)  -W\left(  t\right)  $ are independent of events
prior to $t$ and $\Delta W\sim\mathcal{N}\left(  0,\Delta t\right)  .$ As a
function of $t, $ $W$ is continuous except possibly on a set of measure zero
in $\Omega.$

With this formalism, we can write the stochastic process for $P,$ based on
$\left(  \ref{discP}\right)  ,$ in the limiting form as%
\begin{equation}
dP=G\left(  D/S\right)  Pdt+\sigma G^{\prime}\left(  D/S\right)  \left(
D/S\right)  PdW. \label{contP}%
\end{equation}
Ito's formula applied to the function $\log P$ then yields%
\begin{equation}
d\log P=\left\{  G\left(  D/S\right)  -\frac{1}{2}\left(  \sigma G^{\prime
}\left(  D/S\right)  \left(  D/S\right)  \right)  ^{2}\right\}  dt+\sigma
G^{\prime}\left(  D/S\right)  D/SdW. \label{contLogP}%
\end{equation}
Thus, taking the difference between $\left(  \ref{contP}\right)  $ and
$\left(  \ref{contLogP}\right)  $ yields
\begin{equation}
d\log P=\frac{1}{P}dP+O\left[  \sigma^{2}\right]  . \label{diff}%
\end{equation}
For small $\sigma,$ one can formally neglect the $O\left[  \sigma^{2}\right]
$ terms, and the solutions of the two stochastic equations $\left(
\ref{contP}\right)  $ and $\left(  \ref{contLogP}\right)  $ can be expected to
be close in the appropriate norm. However, we utilize the exact formulation
$\left(  \ref{contP}\right)  $.

Note that equation $\left(  \ref{contP}\right)  $ can be written, using
Condition $G$ in a form that explicitly displays the symmetry between supply
and demand, as%
\begin{equation}
dP=\frac{1}{2}\left\{  G\left(  D/S\right)  -G\left(  S/D\right)  \right\}
Pdt+\sigma\left\{  G^{\prime}\left(  D/S\right)  \left(  D/S\right)
-G^{\prime}\left(  S/D\right)  \left(  S/D\right)  \right\}  PdW
\label{contPsym}%
\end{equation}

The marginal volatility of the price, $P,$ is then defined and calculated (see
Appendix E) as $,$
\begin{equation}
\mathbb{V}_{P}\left(  t\right)  :=\lim_{\Delta t\rightarrow0}\frac{Var\left[
\Delta P\left(  t\right)  \right]  }{\Delta t}=E\left[  \left(  \sigma
PG^{\prime}\left(  D/S\right)  D/S\right)  ^{2}\right]  , \label{vp}%
\end{equation}
where $\Delta P\left(  t\right)  :=P\left(  t+\Delta t\right)  -P\left(
t\right)  .$ We can also write this expression as%
\begin{equation}
\mathbb{V}_{\frac{P}{E\left[  P\right]  }}\left(  t\right)  :=\lim_{\Delta
t\rightarrow0}\frac{1}{\Delta t}Var\left[  \frac{\Delta P\left(  t\right)
}{E\left[  P\left(  t\right)  \right]  }\right]  =E\left[  \left(
\frac{\sigma PG^{\prime}\left(  D/S\right)  D/S}{E\left[  P\right]  }\right)
^{2}\right]  . \label{vpn}%
\end{equation}
Within our formalism, this is the formula that most closely expresses the
marginal volatility of the relative price change. The analogous marginal
volatility for $\log P$ is given by%
\begin{equation}
\mathbb{V}_{\log P}\left(  t\right)  :=\lim_{\Delta t\rightarrow0}%
\frac{Var\left[  \Delta\log P\left(  t\right)  \right]  }{\Delta t}=E\left[
\left(  \sigma G^{\prime}\left(  D/S\right)  D/S\right)  ^{2}\right]  .
\label{vlog}%
\end{equation}
Note that $D$ and $S$ can also be considered as functions of $P$ in which case
$\left(  \ref{contP}\right)  $ becomes a more complicated stochastic
differential equation.

\bigskip

\textbf{5.2. Implications of the supply/demand based stochastic asset dynamics
equation.} The equation $\left(  \ref{contP}\right)  $ has been derived from
supply/demand considerations and the resulting mathematics of a ratio of
normals, while the classical asset price dynamics equation $\left(
\ref{BS}\right)  $ is based on an idealization of empirical observations. The
equation $\left(  \ref{BS}\right)  $ is built on the assumption that the price
change and volatility are independent, while $\left(  \ref{contP}\right)  $
shows that if one considers a supply/demand model that is close in principle
to classical economics, then one is forced into the conclusion that volatility
and price change are closely connected. In particular, $\left(  \ref{vpn}%
\right)  $ has the consequence that when relative prices are changing most
rapidly, the volatility is highest, while it will be lowest at market extrema.
This is consistent with an empirical study of 40 major market tops and
subsequent declines, in which Sornette et. al., 2018 \cite{S} showed that for
two-thirds of the cases, volatility is lowest as the market peaks and higher
as the market rises or falls rapidly. Similarly, Bitcoin exhibits a valley in
volatility near its peak trading price in December 2017 (Caginalp and
Caginalp, 2018 \cite{CC2}).

A key idea in this approach is that the stochastics in prices arises from
random changes in supply and demand near the trading price. In other words, if
one knew how the supply and demand would change, then one could determine how
market makers and arbitrageurs would react, and have an accurate gauge for
price formation.

The approach that leads to $\left(  \ref{contP}\right)  $ has a natural
consequence in that volatility will be greatest when the supply/demand
imbalance is greatest, which is also when the magnitude of the relative price
change is largest. One caveat is that momentum trading, which is not part of
this model, could alter this conclusion. Risk assessment is one of the areas
in which the practical applications of utilizing $\left(  \ref{contP}\right)
$ can lead to better results. Options pricing, e.g., the Black-Scholes
equation, is another.

The analysis we have presented involves only the mathematics of supply and
demand. During market tops and bottoms, there are a number of additional
issues.\ Market bottoms, in particular, feature numerous additional
motivations that are both classical and behavioral. For example, during market
bottoms, one has margin calls, whereby investors who have bought stock
partially with borrowed money are required to sell some stock unless they are
able to provide more cash. In addition there are strategies whereby investors
seek to balance portfolios which require selling as stocks are nearing a
bottom. Behavioral factors such as risk aversion and fear can also play a part
in adding additional volatility near market bottoms. These additional factors
can be built on the general supply/demand model we have presented. In
particular, the expected supply and demand can be determined by differential
equations that are coupled to the price equation $\left(  \ref{contP}\right)
$. With the inclusion of trend based motivations, there is likely to be a more
complex interaction between market extrema and volatility.

Among the practical implications of this analysis is that in measuring
volatility one should select a broad time frame that includes rising, falling,
peaking and bottoming phases of a market in calculating volatility. A
refinement of this concept can be attained by obtaining correlations of
volatility on various time intervals with the price trend on those intervals.

Ultimately, this approach highlights the need for delving deeper into the
motivations of traders and investors so that expected returns and volatility
can be analyzed more comprehensively through data.

Throughout this discussion we are assuming that $D\left(  t\right)  $ and
$S\left(  t\right)  $ are given functions. The analysis is completely general
and allows for any set of motivations or strategies that may determine
$D\left(  t\right)  $ and $S\left(  t\right)  .$ In particular, one can
combine the analysis of this paper with the asset flow equations (see for
example, $\cite{CB}$) in which supply and demand depend upon trend, valuation
and the ratio of available cash to asset. The randomness in the supply and
demand that we have considered in this paper can be incorporated within the
price equation in that system of equations. Furthermore, the supply and demand
can depend on other strategies and behavioral motivations.

\bigskip

\textbf{5.3 Testing using models and empirical data.} If supply and demand are
described by differential equations utilizing microeconomic principles, then a
numerical computation can be utilized to generate a set of prices whereby one
can compute the marginal volatility defined above. In particular, let
$\mathcal{A}:=\left\{  t_{1},t_{2},..t_{M}\right\}  $ be the grid for the
numerical computation on the interval $\left[  0,T\right]  $ with uniform
spacing $\delta t=t_{i}-t_{i-1}.$ Given any $t_{i\ast}\in\mathcal{A}$, we
consider the interval%
\[
\left[  t_{i\ast},t_{i\ast}+\Delta t\right]  =\left[  t_{i\ast},t_{i\ast
+K}\right]  ,
\]
i.e., the interval $\left[  t_{i\ast},t_{i\ast}+\Delta t\right]  $ contains
the $K$ points of the grid $\left\{  t_{i\ast,}t_{i\ast+1},...t_{i\ast
+K}\right\}  $ and $K\delta t=\Delta t.$

Using the relations%
\[
\Delta P=P\left(  t_{i\ast}+\Delta t\right)  -P\left(  t_{i\ast}\right)
=\sum_{j=1}^{K}\left\{  P\left(  t_{i\ast+j\delta t}\right)  -P\left(
t_{i\ast+\left(  j-1\right)  \delta t}\right)  \right\}
\]%
\begin{align*}
Var\left[  \Delta P\right]   &  =Var\left[  \sum_{j=1}^{K}\left\{  P\left(
t_{i\ast+j}\right)  -P\left(  t_{i\ast+\left(  j-1\right)  }\right)  \right\}
\right] \\
&  \tilde{=}KVar\left[  \left\{  P\left(  t_{i\ast+j}\right)  -P\left(
t_{i\ast+\left(  j-1\right)  }\right)  \right\}  \right]  ,
\end{align*}
we treat $\left\{  S_{j}\right\}  =\left\{  P\left(  t_{i\ast+j}\right)
-P\left(  t_{i\ast+\left(  j-1\right)  }\right)  \right\}  $ as a set of
random variables and compute the variance in the usual way, $Var\left[
\left\{  S_{j}\right\}  \right]  =E\left[  S_{j}^{2}\right]  -\left(  E\left[
S_{j}\right]  \right)  ^{2}$ and expectation, $E\left[  P\left(  t\right)
\right]  =K^{-1}\sum_{j=0}^{K-1}P\left(  t_{i\ast+j}\right)  .$

This can be used to compute the approximation to $\left(  \ref{vpn}\right)
$:
\[
\mathbb{V}_{\frac{P}{E\left[  P\right]  }}\left(  t\right)  \tilde{=}\frac
{1}{\Delta t}Var\left[  \frac{\Delta P\left(  t\right)  }{E\left[  P\left(
t\right)  \right]  }\right]  =\frac{Var\left[  \left\{  S_{j}\right\}
\right]  }{\delta t}\frac{1}{E\left[  P\left(  t\right)  \right]  }%
\]
and compared with the exact value.

Given market data in the same form, i.e., a set of values $\mathcal{A}%
:=\left\{  t_{1},t_{2},..t_{M}\right\}  $ and an interval $\left(  t,t+\Delta
t\right)  $ with prices $\left\{  P\left(  t_{i\ast}\right)  ,P\left(
t_{i\ast+1}\right)  ,...,P\left(  t_{i\ast+K}\right)  \right\}  $ one can
compute the marginal volatility above. Using this information one can test
whether the peak of $\mathbb{V}_{P/E\left[  P\right]  }\left(  t\right)  $
occurs as prices change most rapidly, and whether the minimum occurs when
prices are at an extremum. Empirical work by Sornette et. al., 2018 \cite{S}
provides some support for this theoretical conclusion in two-thirds of major
market tops. Similarly, this can similarly be tested in mathematical and
computer generated models.

\bigskip

\textbf{Acknowledgements.} The authors thank two anonymous referees for
raising insightful questions and suggestions for improvement of the manuscript.

\bigskip

\textbf{Conflict of Interest.} The authors declare that they have no conflict
of interest.

\bigskip

\textbf{Appendix A.} \ \textbf{Discussion of market orders versus limit
orders.} \ In most markets such as the NYSE, there are several types of orders
that can be placed. From the perspective of economics, the limit order is
natural; the trader is willing to buy at any price below a set price, and
analogously for sell orders. In a "market order" the trader accepts the
current trading price, and makes the transaction immediately. A market order
appears to be suboptimal from the perspective of basic economics, as the
trader is accepting, without conditions, any prevailing price. Bae et. al.,
2003 \cite{BJ}, however, present the following argument of the advantages of
market orders when conditions are changing rapidly. "Consider a world where
transaction prices move solely in response to information. In this world, the
placement of limit orders is clearly unattractive because they have option
features. A trader who submits a limit buy (sell) order provides the market
with a free put (call) option. When the underlying value of the asset moves
against the trader who submits a limit order, the limit order will be executed
and the trader loses. When the value of an asset moves in favor of the trader,
a limit order will never be executed."

The traders' choice between market and limit orders has been studied in
several papers. Keim and Madhaven 1998, \cite{KM}) provide evidence that
traders such as indexers are likely to use market orders, however, traders who
believe they have information on value tend to use limit orders. In data on 30
stocks from 1990 to 1991, Bae et. al., 2003 \cite{BJ} found that $43.7\%$ of
the trades were market orders. Harris and Hasbrouk, 1996\ \cite{HH} found that
$50\%$ of orders in their data set were market orders. A brokerage
advertisement states: "The vast majority of market orders executed receive a
price better than the nationally published quote." This suggests that the
principal reason for placing a limit order, namely to obtain a better price,
is eliminated. Also, brokers like market orders since they are always
executed, thereby generating a commission, and possibly margin interest,
unlike limit orders which are often not executed.

In addition to market and limit orders, there are several other types of
orders, a frequently used one being "stop-loss," whereby an automatic market
order is placed to sell if the trading price drops below a set level (and
analogously for buy orders). These orders have higher priority than market
orders, which have higher priority than limit orders at the same price. While
the stop-loss seems antithetical to classical economics, it is widely used, as
time is of essence in many cases.

The literature also notes that market orders tend to be preferred, as one
would expect, when the bid/ask spread is small, trading is active, and
volatility is not extremely high (e.g., Bae et. al., 2003 \cite{BJ}). Thus,
market orders can be assumed to be a substantial fraction of the total trades
for stocks with large trading volume.

In determining price changes, the limit orders are less important for several
reasons. First, if the limit price is far from the current trading price, the
order does not have any impact on trading. Second, the limit orders have lower
priority in execution, as noted above. Third, those who place limit orders
generally operate on a longer time scale, so they are less concerned about
short term price changes. For example, an investor who is influenced by value
may feel that the price of $\$20$ for a stock may be a good buy, and would
place an order that is valid for up to $60$ days, with the hope that the stock
will be bought at $\$20$ and will be trading higher in the years to come.

Consequently, in analyzing stock price dynamics, the market orders can be
expected to be the primary factor. The dominance of market orders has the
implication that in the classic method for examining price equilibrium, the
intersection of supply and demand curves becomes degenerate. In the classical
goods model, each consumer has a price below which he is willing to buy, and
analogously for the seller. The underlying assumption for goods and services
is that change is sufficiently gradual that as the supply and demand change,
the trading price shifts to the new equilibrium. However, for stocks, the
supply/demand picture is very heavily weighted near the current trading price,
and prices adjust quickly to new orders so that much of the supply/demand
curves are irrelevant. The supply curve (with price on the horizontal axis,
and quantity on the vertical) can be regarded as a piecewise constant function
with two strictly increasing segments, with the middle plateau representing a
region near the trading price. Likewise, the demand curve would consist of two
strictly decreasing segments with a level segment in between. The classical
intersection is thus replaced by an interval of overlap between supply and
demand. \ To illustrate how these curves would attain this shape, we consider
the hypothetical start of trading, when one can imagine the classical supply
and demand curves are strictly increasing and decreasing, respectively. Once
trading starts, the buyers who had limit buy orders just below the trading
price are willing to increase their bid slightly, as they fear losing out due
to a possible increase in prices. Those who had planned on placing a buy order
above the trading price also see that they can obtain a better price. Although
they would be receiving the market price with the limit order, they can change
to a market order to have higher priority in the trading and avoid missing the
purchase. In this way the smooth intersecting curves evolve into the curves in
which one has plateaus near the trading price.

\bigskip

\bigskip

\textbf{Appendix B.} \textbf{\ Calculation of the second derivative.} We first
state a basic mathematical calculation. Suppose that $r\left(  x\right)  $ and
$s\left(  x\right)  $ are twice continuously differentiable functions whose
derivatives are bounded on bounded subsets of the domain. Then we have the
computation
\[
\frac{\partial^{2}}{\partial x^{2}}H\left(  x\right)  :=-\frac{1}{2}%
\frac{\partial^{2}}{\partial x^{2}}\ \frac{\left(  x+r\left(  x\right)
x^{2}\right)  ^{2}}{c(1+s\left(  x\right)  x)^{2}}
\]%
\begin{align*}
&  =\allowbreak-\frac{1}{c\left(  xs\left(  x\right)  +1\right)  ^{2}}\left(
2xr\left(  x\right)  +x^{2}\frac{\partial r\left(  x\right)  }{\partial
x}+1\right)  ^{2}\\
&  -\allowbreak\frac{3}{c}x^{2}\left(  s\left(  x\right)  +x\frac{\partial
s\left(  x\right)  }{\partial x}\right)  ^{2}\frac{\left(  xr\left(  x\right)
+1\right)  ^{2}}{\left(  xs\left(  x\right)  +1\right)  ^{4}}\\
&  -\allowbreak\frac{1}{c\left(  xs\left(  x\right)  +1\right)  ^{2}}\left(
x+x^{2}r\left(  x\right)  \right)  \left(  2r\left(  x\right)  +4x\frac
{\partial r\left(  x\right)  }{\partial x}+x^{2}\frac{\partial^{2}r\left(
x\right)  }{\partial x\partial x}\right) \\
&  -\allowbreak\frac{1}{2c}x^{2}\frac{\left(  xr\left(  x\right)  +1\right)
^{2}}{\left(  xs\left(  x\right)  +1\right)  ^{3}}\left(  -4\frac{\partial
s\left(  x\right)  }{\partial x}-2x\frac{\partial^{2}s\left(  x\right)
}{\partial x\partial x}\right) \\
&  -\allowbreak\frac{1}{2c\left(  xs\left(  x\right)  +1\right)  ^{3}}\left(
x+x^{2}r\left(  x\right)  \right)  \left(  -8s\left(  x\right)  -8x\frac
{\partial s\left(  x\right)  }{\partial x}\right)  \allowbreak\left(
2xr\left(  x\right)  +x^{2}\frac{\partial r\left(  x\right)  }{\partial
x}+1\right)  .
\end{align*}
Evaluating this expression at $x=0$ and using the boundedness properties of
$r\left(  x\right)  $ and $s\left(  x\right)  ,$ one has then%

\[
\frac{\partial^{2}}{\partial x^{2}}H\left(  0\right)  =-\frac{1}{c}.
\]

We apply this proposition to $\left(  \ref{h1}\right)  $ with
\begin{align*}
r\left(  z\right)   &  :=\frac{1}{2\Delta t}G^{\prime}\left(  \frac{D}%
{S}\right)  \left(  G^{-1}\right)  ^{\prime\prime}\left(  \zeta_{2}\left(
z\right)  \right)  ,~\ \ c:=2\left(  \frac{D}{S}G^{\prime}\left(  \frac{D}%
{S}\right)  \right)  ^{2}\\
s\left(  z\right)   &  :=\frac{1}{2\frac{D}{S}G^{\prime}\left(  \eta
_{1}\right)  \Delta t}=\frac{\left(  G^{-1}\right)  ^{\prime}\left(  \zeta
_{1}\left(  z\right)  \right)  }{2\Delta t\frac{D}{S}}.
\end{align*}
Recall that, from the original definitions, $\zeta_{1}$ and $\zeta_{2}$ are
functions of $z,$ as they are values between $y/\Delta t-y_{0}/\Delta t$ while
$z:=y-G\left(  D/S\right)  \Delta t.$ Thus $y\rightarrow G\left(  D/S\right)
\Delta t$ as $z\rightarrow0.$ For example, since $y_{0}/\Delta t=G\left(
D/S\right)  $ as $z\rightarrow0,$ one has $\zeta_{1,2}\rightarrow G\left(
D/S\right)  \Delta t.$ Clearly, then, on a bounded interval containing
$y_{0}/\Delta t$, the first four derivatives of $G^{-1}$ are bounded so that
$r^{\prime\prime}\left(  z\right)  $ and $s^{\prime\prime}\left(  z\right)  $
are bounded. We have then%
\[
\frac{d^{2}h\left(  0\right)  }{dz^{2}}=-\frac{1}{\left(  \frac{D}{S}%
G^{\prime}\left(  \frac{D}{S}\right)  \right)  ^{2}}.
\]
$\allowbreak$

\bigskip

\bigskip

\textbf{Appendix C. An alternative scaling.} Another way to choose the small
parameter starts with the same expressions for $f_{3}$, namely $\left(
\ref{f3}\right)  $, in terms of $E$ and $B.$ Now using, instead of $z,$ the
variable $w:=\left(  y-G\left(  D/S\right)  \Delta t\right)  /\Delta t$ and
defining $R\left(  y\right)  =:\hat{R}\left(  w\right)  $ leads to%
\[
I_{2}=\int_{-\infty}^{\infty}R\left(  y\right)  f_{3}\left(  y\right)
dy=\int_{-\infty}^{\infty}\hat{R}\left(  w\right)  \hat{f}_{3}\left(
w\right)  \Delta tdw,
\]%
\[
\hat{f}_{3}\left(  w\right)  \Delta tdw=\frac{\exp\left\{  -\frac{1}{2}%
\frac{\left(  \frac{1}{G^{\prime}\left(  \frac{D}{S}\right)  }w+\frac{\left(
G^{-1}\right)  ^{\prime\prime}\left(  \zeta_{2}\right)  }{2}w^{2}\right)
^{2}}{\frac{\sigma^{2}}{\Delta t}\frac{1}{4}\left(  2\frac{D}{S}+\frac
{1}{G^{\prime}\left(  \eta_{1}\right)  }w\right)  ^{2}}\right\}  dw}%
{\sqrt{2\pi}\frac{\sigma}{\left(  \Delta t\right)  ^{1/2}}G^{\prime}\left(
G^{-1}\left(  \frac{y}{\Delta t}\right)  \right)  \frac{D}{S}\left(
1+\frac{1}{2\frac{D}{S}G^{\prime}\left(  \eta_{1}\right)  }w\right)  ^{2}}.
\]
At this point the integral $I_{2}$ has the same form as the previous scaling.
Upon defining $\alpha:=\Delta t/\sigma^{2},$
\[
h_{1}\left(  w\right)  :=-\frac{1}{2}\frac{\left(  \frac{1}{G^{\prime}\left(
\frac{D}{S}\right)  }w+\frac{\left(  G^{-1}\right)  ^{\prime\prime}\left(
\zeta_{2}\right)  }{2}w^{2}\right)  ^{2}}{\frac{1}{4}\left(  2\frac{D}%
{S}+\frac{1}{G^{\prime}\left(  \eta_{1}\right)  }w\right)  ^{2}}%
\]%
\[
u_{2}\left(  w\right)  :=\frac{\hat{R}\left(  w\right)  }{\sqrt{2\pi}%
G^{\prime}\left(  G^{-1}\left(  \frac{y}{\Delta t}\right)  \right)  \frac
{D}{S}\left(  1+\frac{1}{2\frac{D}{S}G^{\prime}\left(  \eta_{1}\right)
}w\right)  ^{2}}%
\]
the integral can be expressed as%
\[
I_{2}=\alpha^{1/2}\int_{-\infty}^{\infty}u_{2}\left(  w\right)  e^{\alpha
h_{1}\left(  w\right)  }dw.
\]
The standard result above then yields
\begin{align}
I_{2}  &  =\alpha^{1/2}\left\{  u\left(  0\right)  \left(  \frac{-2\pi}{\alpha
h_{1}^{\prime\prime}\left(  0\right)  }\right)  ^{1/2}+O\left(  \alpha
^{-3/2}\right)  \right\} \nonumber\\
&  =\frac{\hat{R}\left(  0\right)  }{\sqrt{2\pi}G^{\prime}\left(  D/S\right)
\frac{D}{S}}\frac{\sqrt{2\pi}}{h_{1}^{\prime\prime}\left(  0\right)
}+O\left(  \frac{\sigma^{2}}{\Delta t}\right) \nonumber\\
&  =R\left(  D/S\right)  +O\left(  \frac{\sigma^{2}}{\Delta t}\right)  .
\label{I2}%
\end{align}

The same result holds for $I_{1}$ as we show below:%
\[
I_{2}=\int_{-\infty}^{\infty}R\left(  y\right)  f_{3}^{\left(  N\right)
}\left(  y\right)  dy=\int_{-\infty}^{\infty}\hat{R}\left(  w\right)
f_{3}^{\left(  N\right)  }\left(  w\right)  \Delta tdw,
\]%
\[
\hat{f}_{3}\left(  w\right)  \Delta tdw=\frac{\exp\left\{  -\frac{1}{2}%
\frac{\left(  \frac{1}{G^{\prime}\left(  \frac{D}{S}\right)  }w\right)  ^{2}%
}{\frac{\sigma^{2}}{\Delta t}\left(  \frac{D}{S}\right)  ^{2}}\right\}
dw}{\sqrt{2\pi}\frac{\sigma}{\left(  \Delta t\right)  ^{1/2}}G^{\prime}\left(
\frac{D}{S}\right)  \frac{D}{S}}.
\]
Defining
\[
h_{0}\left(  w\right)  :=-\frac{1}{2}\frac{\left(  \frac{1}{G^{\prime}\left(
\frac{D}{S}\right)  }w\right)  ^{2}}{\left(  \frac{D}{S}\right)  ^{2}}%
\]
so $h_{0}\left(  0\right)  =0$ and $h_{0}^{^{\prime\prime}}\left(  w\right)
=-\frac{1}{\left(  G^{\prime}\left(  \frac{D}{S}\right)  \frac{D}{S}\right)
^{2}}$ and
\[
u_{0}\left(  w\right)  :=\frac{\hat{R}\left(  w\right)  }{\sqrt{2\pi}%
G^{\prime}\left(  \frac{D}{S}\right)  \frac{D}{S}}%
\]
yields the calculation
\begin{align}
I_{1}  &  =\alpha^{1/2}\left\{  u_{0}\left(  0\right)  \left(  \frac{-2\pi
}{\alpha h_{0}^{\prime\prime}\left(  0\right)  }\right)  ^{1/2}+O\left(
\alpha^{-3/2}\right)  \right\} \nonumber\\
&  =\alpha^{1/2}\left\{  \frac{\hat{R}\left(  0\right)  }{\sqrt{2\pi}%
G^{\prime}\left(  \frac{D}{S}\right)  \frac{D}{S}}\left(  \frac{2\pi\left(
G^{\prime}\left(  \frac{D}{S}\right)  \frac{D}{S}\right)  ^{2}}{\alpha
}\right)  ^{1/2}+O\left(  \alpha^{-3/2}\right)  \right\} \nonumber\\
&  =R\left(  \frac{D}{S}\right)  +O\left(  \frac{\sigma^{2}}{\Delta t}\right)
. \label{I1}%
\end{align}

Comparing $\left(  \ref{I1}\right)  $ and $\left(  \ref{I2}\right)  $, we see
that the difference between $I_{1}$ and $I_{2}$ is at most $C\sigma^{2}/\Delta
t$ .

\bigskip

\bigskip

\textbf{Appendix D. Theorems on convergence.} \ After two definitions, we
state a theorem from probability (see, e.g., Billingsley 2008 \cite{BI} \ p.
352) that is utilized in the derivation.

\bigskip

\textbf{Definition}. Let $X_{n}$ and $X$ be random variables with
distributions $F_{n}$ and $F.$ We say that $F_{n}$ converges weakly to $F$
(denoted $F_{n}\Longrightarrow F$) if%
\[
\lim_{n\rightarrow\infty}F_{n}\left(  x\right)  =F\left(  x\right)
\]
for every point of continuity $x.$ We say that $X_{n}$ converges in
distribution, denoted $X_{n}\Longrightarrow X$, if $F_{n}\Longrightarrow F$.

\bigskip

\textbf{Definition}. If $\mu_{n}$ and $\mu$ are the probability measures
corresponding to $F_{n}$ and $F$ then we write $\mu_{n}\Longrightarrow\mu$ if%
\[
\lim_{n\rightarrow\infty}\mu_{n}\left(  A\right)  =\mu\left(  A\right)
\]
for every set of the form $A:=(-\infty,x].$

\bigskip

\textbf{Theorem D.1}. The following are equivalent:

$\left(  i\right)  $ $\mu_{n}\Longrightarrow\mu$, $\left(  ii\right)  $
$\lim_{n\rightarrow\infty}\mu_{n}\left(  A\right)  =\mu\left(  A\right)  $ for
every $\mu-$continuity set A.

$\left(  iii\right)  $ $\int gd\mu_{n}\rightarrow\int gd\mu$ \ for every
bounded, continuous real function $g.$

\bigskip

\textbf{Appendix E. \ Calculation of volatility.} Let $\mathcal{F}_{t}$ be a
Brownian filtration (see Schuss, 2009 \cite{SC} p. 66 ) so that $\mathcal{F}%
_{t}$ adapted stochastic processes $f\left(  t;\omega\right)  $ are
independent of increments of the Brownian motion, $W\left(  t;\omega\right)
,$ that are prior to $t$. Let $H_{2}\left[  0,T\right]  $ consist of
$\mathcal{F}_{t}$ adapted stochastic processes $f\left(  t;\omega\right)  $ on
the interval $\left[  0,T\right]  $ such that
\[
\int_{0}^{T}Ef^{2}\left(  s;\omega\right)  ds<\infty.
\]
Now let $f\left(  t;\omega\right)  $ $g\left(  t;\omega\right)  $ be
stochastic processes in $H_{2}\left[  0,T\right]  $ such that for some
positive constant $C$ one has
\begin{equation}
\sup_{t\in\left[  0,T\right]  }\left\{  Ef^{2}\left(  t;\omega\right)
,Eg^{2}\left(  t;\omega\right)  \right\}  <C. \label{sup bound}%
\end{equation}
Also, assume that $f$ and $g$ are continuous in $t$ except possibly on a set
of measure zero in $\Omega$. We now consider a general stochastic process
$Z\left(  t;\omega\right)  $ defined by%
\begin{equation}
dZ=fdt+gdW \label{z}%
\end{equation}
which is equivalent to writing (suppressing the $\omega$ variable)%
\begin{equation}
Z\left(  t\right)  -Z\left(  t_{0}\right)  =\int_{t_{0}}^{t}f\left(  s\right)
ds+\int_{t_{0}}^{t}g\left(  s\right)  dW\left(  s\right)  . \label{zi}%
\end{equation}

\bigskip

\textbf{Theorem E.1.} If the stochastic processes $f$ and $g$ satisfy $\left(
\ref{sup bound}\right)  $, $Z$ is a stochastic process defined by $\left(
\ref{zi}\right)  $, and $\Delta Z\left(  t\right)  :=Z\left(  t+\Delta
t\right)  -Z\left(  t\right)  ,$ then
\begin{equation}
Var\left[  \Delta Z\right]  =Var\left[  \int_{t}^{t+\Delta t}fds\right]
+2E\left[  \int_{t}^{t+\Delta t}fds\int_{t}^{t+\Delta t}gdW\right]  +\int
_{t}^{t+\Delta t}Eg^{2}ds \label{var}%
\end{equation}
and the marginal volatility $\mathbb{V}\left(  t\right)  $ satisfies
\begin{equation}
\mathbb{V}\left(  t\right)  :=\lim_{\Delta t\rightarrow0}\frac{Var\left[
\Delta Z\left(  t\right)  \right]  }{\Delta t}=Eg^{2}\left(  t\right)  .
\label{vol}%
\end{equation}

\textbf{Proof.} The first identity, $\left(  \ref{var}\right)  ,$ follows from
squaring $\left(  \ref{zi}\right)  $, \ and noting that
\[
E\int_{t}^{t+\Delta t}g\left(  s\right)  dW\left(  s\right)  =0
\]
from the assumption that $g$ is $\mathcal{F}_{t}$ adapted, and using a basic
property of Ito integrals (see e.g., Schuss, 2009 \cite{SC}, p. 68):
\[
E\left[  \left(  \int_{t}^{t+\Delta t}gdW\right)  ^{2}\right]  =\int
_{t}^{t+\Delta t}Eg^{2}ds
\]
To obtain the second identity we establish the following two bounds using the
Cauchy-Schwarz inequality for integrals, and then for expectations. Note that
from $\left(  \ref{sup bound}\right)  $ one has
\[
\left(  \int_{t}^{t+\Delta t}fds\right)  ^{2}\leq\left(  \int_{t}^{t+\Delta
t}1^{2}ds\right)  \left(  \int_{t}^{t+\Delta t}f^{2}ds\right)  ,
\]%
\[
E\left[  \left(  \int_{t}^{t+\Delta t}fds\right)  ^{2}\right]  \leq\Delta
t\int_{t}^{t+\Delta t}Ef^{2}ds\leq C\left(  \Delta t\right)  ^{2}%
\]
upon taking expectations. Now, using Cauchy-Schwarz for expectations on the
middle term in $\left(  \ref{var}\right)  $ yields%
\begin{align*}
\left\vert E\left[  \int_{t}^{t+\Delta t}fds\int_{t}^{t+\Delta t}gdW\right]
\right\vert  &  \leq E\left[  \left(  \int_{t}^{t+\Delta t}fds\right)
^{2}\right]  ^{1/2}E\left[  \left(  \int_{t}^{t+\Delta t}gdW\right)
^{2}\right]  ^{1/2}\\
&  \leq C\Delta t\left(  \int_{t}^{t+\Delta t}Eg^{2}ds\right)  ^{1/2}\leq
C^{2}\left(  \Delta t\right)  ^{3/2}.
\end{align*}
Upon dividing by $\Delta t$ and taking the limit of $\left(  \ref{var}\right)
$ one has from $\left(  \ref{vol}\right)  $
\[
\mathbb{V}\left(  t\right)  =\lim_{\Delta t\rightarrow0}\int_{t}^{t+\Delta
t}Eg^{2}\left(  s;\omega\right)  ds=Eg^{2}\left(  t;\omega\right)
\]
since $g$ and $g^{2}$ are continuous in $t$ except possibly on a set of
measure zero in $\Omega.$ Thus, $\left(  Eg^{2}\right)  \left(  t\right)  $ is
a continuous function of $t.$ $///$

If $f$ and $g$ are functions in $\left(  \ref{zi}\right)  $ also\ depend on
$Z, $ i.e., $f\left(  Z,t;\omega\right)  ,$ and initial conditions $Z\left(
0\right)  :=Z_{0}$ are imposed, we have a stochastic differential equation
(SDE). We say that $Z$ is a solution in the Ito sense if $Z\left(
t;\omega\right)  \in H_{2}\left[  0,T\right]  $ for all $T>0$ and $\left(
\ref{zi}\right)  $ holds for almost all $\omega\in\Omega$. Also $f$ and $g$
satisfy a uniform Lipschitz condition if%
\begin{equation}
\left\vert f\left(  Z,t;\omega\right)  -f\left(  Y,t;\omega\right)
\right\vert +\left\vert g\left(  Z,t;\omega\right)  -g\left(  Y,t;\omega
\right)  \right\vert \leq K\left\vert Z-Y\right\vert \label{lc}%
\end{equation}
for some $K>0$ and all $Y,Z\in\mathbb{R}$.

The stochastic differential equation $\left(  \ref{zi}\right)  $ has a unique
solution in the Ito sense if $f$ and $g$ satisfy $\left(  \ref{lc}\right)  $
for all $t$ and for almost all $\omega\in\Omega.$ The trajectories $Z\left(
t;\omega\right)  $ are then continuous for almost all $\omega\in\Omega$ (see
e.g., Schuss, 2009 \cite{SC}, p. 94).

$\left(  i\right)  $ Applying these results first to $Z:=\log P$ and
$f:=G\left(  D/S\right)  ,$ $g:=\sigma G^{\prime}\left(  D/S\right)  D/S$,
i.e., the SDE
\[
d\log P=G\left(  D/S\right)  dt+\sigma G^{\prime}\left(  D/S\right)  D/SdW,
\]
one obtains from the theorem above, the result, with $\Delta\log P\left(
t\right)  :=\log P\left(  t+\Delta t\right)  -\log P\left(  t\right)  ,$%
\begin{equation}
\mathbb{V}_{\log P}\left(  t\right)  :=\lim_{\Delta t\rightarrow0}%
\frac{Var\left[  \Delta\log P\left(  t\right)  \right]  }{\Delta t}=E\left[
\left(  \sigma G^{\prime}\left(  D/S\right)  D/S\right)  ^{2}\right]  .
\label{vlog1}%
\end{equation}

$\left(  ii\right)  $ Next, applying these results to $Z:=P$ with
$f:=PG\left(  D/S\right)  ,$ $g:=\sigma PG^{\prime}\left(  D/S\right)  D/S$,
i.e., the SDE%
\[
dP=PG\left(  D/S\right)  dt+\sigma PG^{\prime}\left(  D/S\right)  D/SdW
\]
yields, with $\Delta P\left(  t\right)  :=P\left(  t+\Delta t\right)
-P\left(  t\right)  ,$ the marginal volatility,
\begin{equation}
\mathbb{V}_{P}\left(  t\right)  :=\lim_{\Delta t\rightarrow0}\frac{Var\left[
\Delta P\left(  t\right)  \right]  }{\Delta t}=E\left[  \left(  \sigma
PG^{\prime}\left(  D/S\right)  D/S\right)  ^{2}\right]  , \label{vp1}%
\end{equation}
which we can also write as%
\begin{equation}
\mathbb{V}_{\frac{P}{E\left[  P\left(  t\right)  \right]  }}\left(  t\right)
:=\lim_{\Delta t\rightarrow0}\frac{1}{\Delta t}Var\left[  \frac{\Delta
P\left(  t\right)  }{E\left[  P\left(  t\right)  \right]  }\right]  =E\left[
\left(  \frac{\sigma PG^{\prime}\left(  D/S\right)  D/S}{E\left[  P\left(
t\right)  \right]  }\right)  ^{2}\right]  .\ \ \ \ \ /// \label{vpn1}%
\end{equation}

\end{document}